\documentclass[a4paper]{spie}  
\usepackage[]{graphicx}
\usepackage[draft]{hyperref}

\title{2.23\,GHz gating InGaAs/InP single-photon avalanche diode for quantum key distribution}

\author{Jun~Zhang$^{*}$\supit{a}, Patrick~Eraerds\supit{a}, Nino~Walenta\supit{a}, Claudio~Barreiro\supit{a}, Rob~Thew\supit{a}, and Hugo~Zbinden\supit{a}
\skiplinehalf
\supit{a}Group of Applied Physics, University of Geneva, 1211 Geneva 4, Switzerland \\}
\authorinfo{Further author information: (Send correspondence to Jun~Zhang)\\E-mail: Jun.Zhang@unige.ch;
Telephone: +41 (0)22 379 68 82; Fax: +41 (0)22 379 39 80 \\}

\begin{document}
\maketitle

\begin{abstract}
We implement an InGaAs/InP single-photon avalanche diode (SPAD) for single-photon detection
with the fastest gating frequency reported so far, of 2.23\,GHz, which approaches
the limit given by the bandwidth of the SPAD - 2.5\,GHz. We propose a useful way
to characterize the afterpulsing distribution for rapid gating that allows for
easy comparison with conventional gating regimes. We compare the performance
of this rapid gating scheme with free-running detector and superconducting
single-photon detector (SSPD) for the coherent one-way quantum key
distribution (QKD) protocol. The rapid gating system is well suited for both
high-rate and long-distance QKD applications, in which Mbps key rates
can be achieved for distances less than 40\,km with 50\,ns deadtime and
the maximum distance is limited to $\sim$190\,km with 5\,$\mu$s deadtime.
These results illustrate that the afterpulsing is no longer a limiting factor for QKD.
\end{abstract}


\section{INTRODUCTION}
\label{sec:intro}

Near-infrared single-photon detection is one of the key components for diverse applications, e.g.,
quantum key distribution (QKD) \cite{GRTZ02} or optical time domain reflectometry \cite{ELZZG10}.
InGaAs/InP SPADs working in the Geiger mode can provide a practical and
reliable solution \cite{RGZG98,Itzler07}. The quenching electronics \cite{Cova96} operating on these devices
has been extensively investigated for more than two decades and has provided the most significant
performance improvements.
The avalanche amplitude is highly dependent on the excess bias voltage and the duration
of the avalanche \cite{ZTGGZ09}. Larger amplitudes are easier to discriminate, but
more carriers are trapped by the defects in the multiplication layer \cite{Itzler07}. These carriers can be
subsequently released and create undesired avalanches, called afterpulses.
The population of trapped carriers exponentially decays in time, and
increasing temperatures can accelerate the depopulation process and hence decrease the afterpulsing.
Generally, the afterpulsing effect is the most limiting factor for SPAD performance.
Apart from increasing deadtime or heating the diode, there are other approaches to decrease the afterpulsing,
e.g., speeding up the quenching time by integrating the quenching electronics into a single chip \cite{TSGZR07,ZTGGZ09},
or utilizing rapid gating with ultrashort gating durations \cite{NSI06,NAI09,YKSS07,DDYSBS09,ZTBZ09}.

In general, in rapid gating systems, the gating repetition frequency ($f_{g}$) can reach
the GHz level and the effective gating width ($t_{g}$) is usually below 300\,ps. Therefore, the
number of carriers created during an avalanche and thus the afterpulsing
is significantly reduced but at the same time the avalanche amplitude becomes
quite small, i.e., a few mV in general \cite{NSI06,NAI09,YKSS07,DDYSBS09,ZTBZ09}.
The essence of rapid gating is to then extract faint avalanches while
maintaining a sufficient signal-noise ratio (SNR) between
avalanche signals and capacitive response signals. So far, there are two methods to
implement rapid gating, i.e., sine wave gating and filtering \cite{NSI06,NAI09},
self-differencing \cite{YKSS07,DDYSBS09} as well as a hybrid approach combining
the above techniques \cite{ZTBZ09}. For high-speed synchronous single-photon
detection, $f_{g}$ is a crucial parameter. Firstly, the value of $f_{g}$
determines the operation speed of the whole system in applications. Secondly,
increasing $f_{g}$ correspondingly decreases $t_{g}$ and thus the afterpulsing, for a fixed
excess bias ($V_{e}$). Most InGaAs/InP SPADs used for single-photon detection were
originally designed for 2.5\,Gbps classical optical communication. Considering the frequency
response limits of these photodiodes and the final SNRs, the limit of $f_{g}$ for rapid gating
could be presumed to be around 2.5\,GHz.

In this Letter we report a rapid gating scheme with $f_{g}$=2.23\,GHz
that approaches the above mentioned frequency limit with detection
efficiencies over 10\,\%. We also illustrate a simple method to
characterize the afterpulsing distribution in time that allows for easy comparison
with conventional gating SPADs. Finally, we simulate the performance difference
between rapid gating and free-running SPADs for QKD applications.

\section{THE EXPERIMENT}
\label{sec:exp}

\begin{figure}[]
\centering
\includegraphics[width=12 cm]{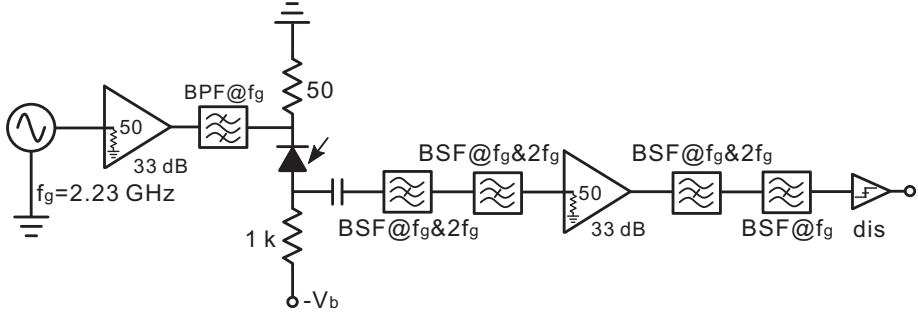}\\
\caption{The experimental setup. BPF: band-pass filter; BSF: band-stop filter; $f_{g}$: gating frequency;
dis: discriminator; $-V_{b}$: negative dc bias; @: the center frequency.}
\label{scheme}
\end{figure}

\begin{figure}[t]
\centering
\includegraphics[width=12 cm]{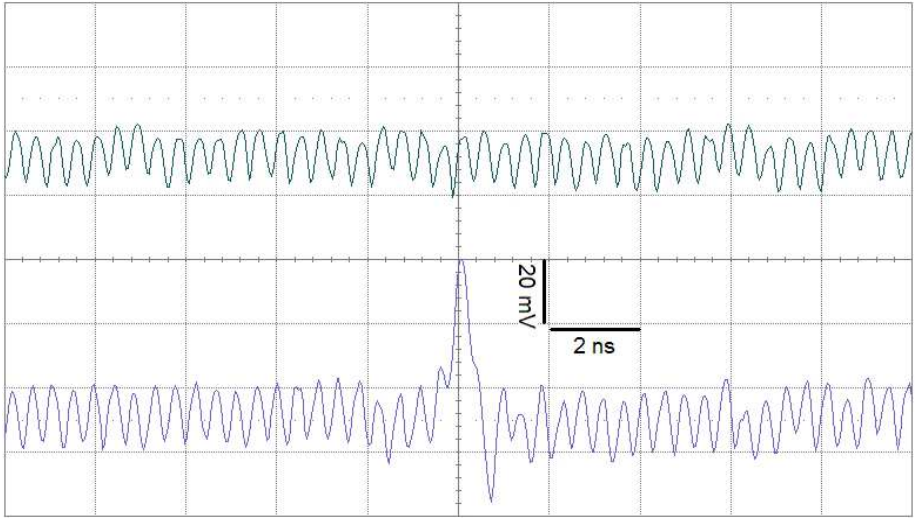}\\
\caption{Typical capacitive response signal without avalanche (top) and
avalanche signal (bottom) observed on oscilloscope after four BSFs and
the amplifier. The horizontal and vertical units are 2\,ns/division
and 20\,mV/division, respectively.}
\label{wave}
\end{figure}

The entire detection scheme is depicted in Figure \ref{scheme}. In our experiment,
we employ the method of sine gating and filtering. The original
sine wave signals from the generator (Agilent E4433B) pass through a 33\,dB amplifier
(Mini-Circuits ZHL-42W) and a band-pass filter (BPF) to produce gates with peak-peak
amplitude ($V_{pp}$) of $\sim$ 7.5\,V. The output signals from the SPAD are processed by four band-stop
filters (BSFs) and another 33\,dB amplifier to extract the avalanche signals.
The BSFs suppress the capacitive response signals induced by the SPAD at the
fundamental frequency $f_{g}$ and also the second harmonic 2$f_{g}$.
All the BSFs and the BPF are designed and fabricated using microstrips (FR4 substrate,
dielectric constant $\varepsilon$=4.5, height $H$=1.43\,mm, thickness $T$=35\,$\mu$m, and $Z_{in}$=$Z_{out}\equiv$50\,$\Omega$).
The final response and avalanche signals in front of the discriminator (dis) are captured
by an oscilloscope (LeCroy WaveMaster 8600A, 6\,GHz bandwidth and 20\,GS/s), see
Figure \ref{wave}. We can deduce that the typical amplitudes of response and avalanche signals without
the amplifier should be $\sim$0.5\,mV and 1\,mV, respectively. The avalanche amplitude
depends on many factors such as the actual gating duration, the excess bias $V_{e}$, the variation of
the multiplication gain, the absorbed photon number and so on.

\section{RESULTS AND DISCUSSIONS}
\label{sec:res}

\begin{figure}[]
\centering
\includegraphics[width=11 cm]{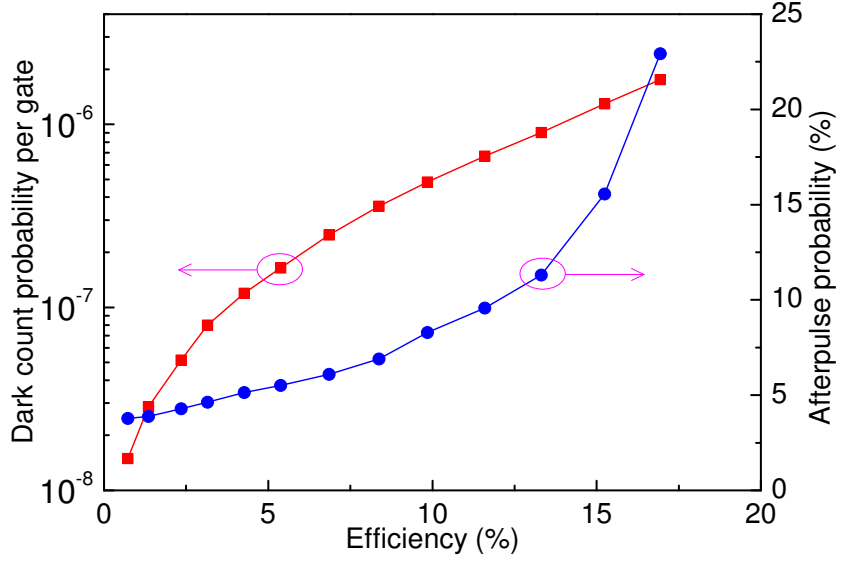}\\
\caption{Plot of dark count probability per gate and afterpulse probability as a function of
efficiency at T=-40\,$^\circ$C and a mean photon number per laser pulse of 0.1.}
\label{eff}
\end{figure}

\begin{figure}[]
\centering
\includegraphics[width=11 cm]{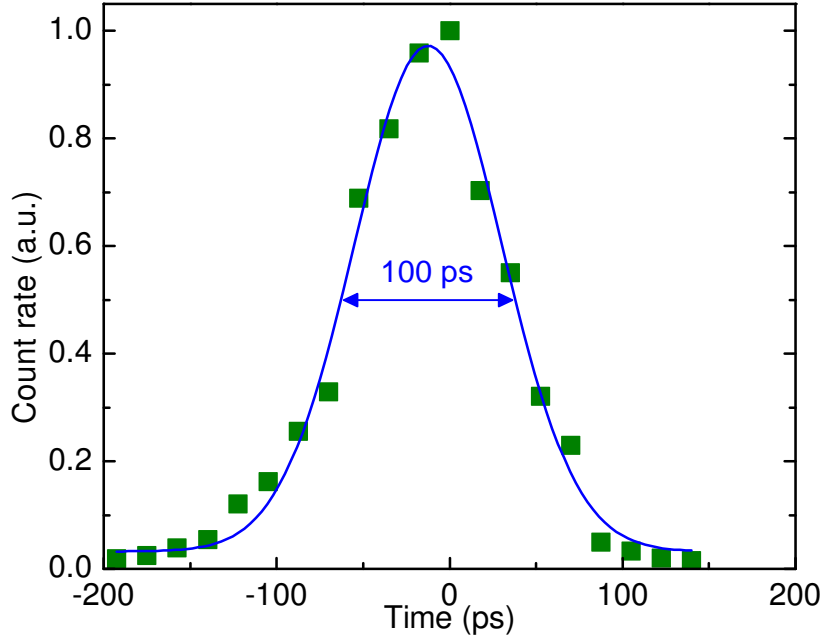}\\
\caption{Count rate distribution obtained by delaying the position of the laser pulse with
respect to the detection gate under the conditions of 10\,\% efficiency and T=-40\,$^\circ$C.
The effective gating width ($t_{g}$) is 100\,ps.}
\label{gate}
\end{figure}

To characterize the performance of the system, we use the same SPAD (\#1 SPAD) and the same calibration methods as used
in Ref. \cite{ZTBZ09}. The 10\,MHz synchronous output from the generator drives an ultrashort
laser diode in the telecom regime (1550\,nm) with $\sim$30\,ps width (PicoQuant PDL 800-B), which is
attenuated down to the single-photon level per pulse. We apply a 10\,ns ``deadtime'' \cite{ZTBZ09} to the discriminator output, which
means that once a detection is recorded all signals in the following 10\,ns are ignored. The
photon detections are counted by the coincidences between the laser pulses and the detections,
while all the remaining detections are attributed to dark counts and afterpulses. The calibrated
results are shown in Figure \ref{eff}. The efficiency $\eta$ is calculated as \cite{ZTBZ09}
\begin{equation}
\label{}
\eta=\frac {1} {\mu}\cdot \ln \frac {1-P_{dc}}{1-P_{de}},
\end{equation}
where $\mu$ is the mean photon number per laser
pulse, $P_{dc}$ is the dark count probability per gate and $P_{de}$ is the photon detection probability
per laser pulse. When $\eta$=10\,\%, the afterpulse probability ($P_{ap}$) is 8.3\,\% at T=-40\,$^\circ$C,
equivalent to $\sim$4$\times$10$^{-5}$ ns$^{-1}$ calculated according to Eqn. 3 in Ref. \cite{ZTBZ09}.
$P_{dc}$ is 4.8$\times$10$^{-7}$ per gate, equivalent to 4.8$\times$10$^{-6}$ ns$^{-1}$
since $t_{g}$=100\,ps, see Figure \ref{gate}. At a fixed excess bias $V_{e}$, the
value of $t_{g}$ depends on the sine gate amplitude $V_{pp}$. Larger $V_{pp}$ corresponds to smaller $t_{g}$,
which can suppress the afterpulsing effect but degrades the SNR and increases the processing
difficulty for the back-end electronics.

\begin{figure}[]
\centering
\includegraphics[width=11 cm]{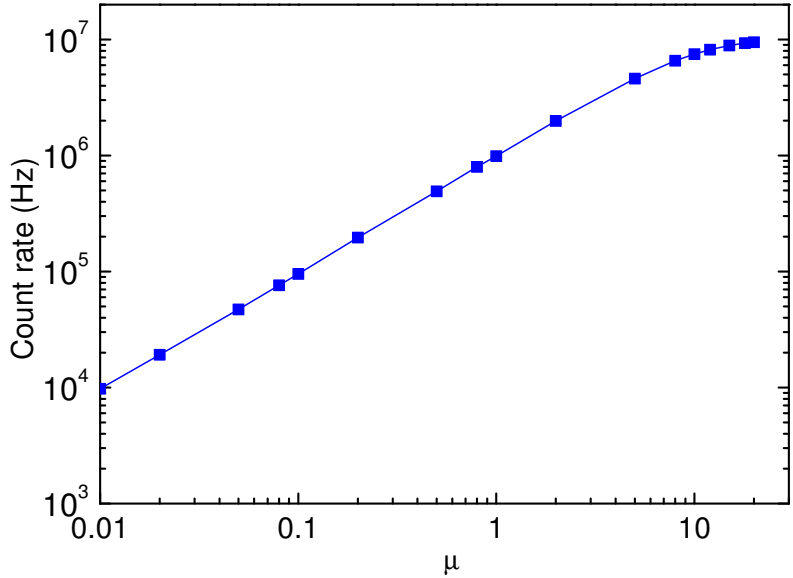}\\
\caption{Count rates versus the mean photon number per laser pulse at 10\,\% efficiency. The saturated
count rate reaches $\sim$10\,MHz, i.e., the laser frequency.}
\label{rate}
\end{figure}

\begin{figure}[]
\centering
\includegraphics[width=11 cm]{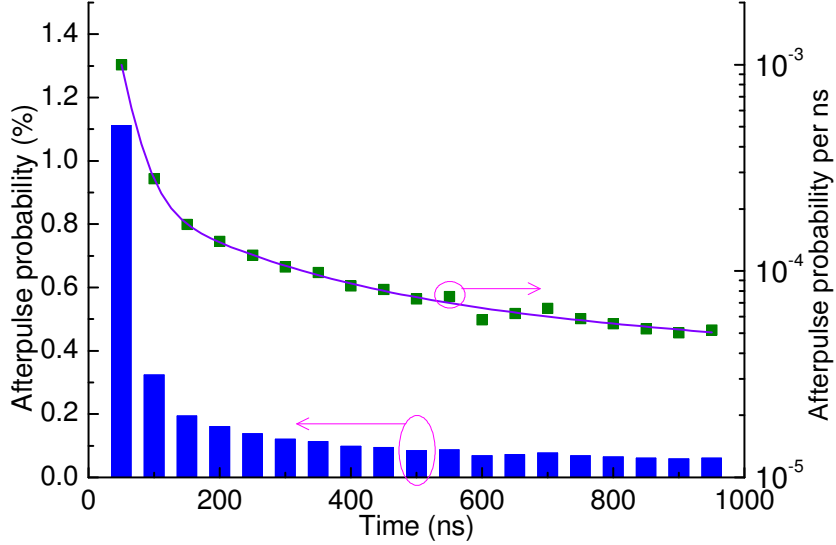}\\
\caption{The afterpulsing distribution in a 1\,$\mu$s range with a laser pulse frequency of 500\,kHz
at 10\,\% efficiency and T=-40\,$^\circ$C. The histogram (left axis) is measured by
scanning the position of a 50\,ns coincidence window.
The corresponding normalized parameter $P_{ap}$[ns$^{-1}$] (right axis) is plotted and fitted.}
\label{dis}
\end{figure}

Let us compare the normalized parameters of the SPAD to the results in Ref. \cite{ZTBZ09},
where at $f_{g}$=921\,MHz, T=-30\,$^{\circ}$C and $\eta$=9.3\,\%, $P_{dc}$ per ns ($P_{dc}$[ns$^{-1}$])
and $P_{ap}$ per ns ($P_{ap}$[ns$^{-1}$]) are 2.8$\times10^{-6}$ ns$^{-1}$ and 1.6$\times10^{-4}$ ns$^{-1}$ \cite{ZTBZ09}.
The values of $P_{dc}$[ns$^{-1}$] are at the same levels and the small difference is probably
due to the $V_{pp}$ difference. $P_{ap}$[ns$^{-1}$] in our experiment with even cooler
temperature is only 1/4 of the value in Ref. \cite{ZTBZ09}. The main reason for the improvement
is attributed to the smaller $t_{g}$.

Figure \ref{rate} illustrates the count rate characteristic as a function of mean photon number per laser pulse.
As $\mu$ rises the count rate linearly increases when $\mu<$10, and finally the count rate is saturated to
$\sim$10\,MHz, which is the same as the laser frequency. Since this frequency was fixed inside the generator
and the maximum laser driver frequency was also limited, we could not test the theoretically maximum count rate
of SPAD, i.e., $\sim$100\,MHz given by the 10\,ns ``deadtime'' setting.

We also use the coincidence method \cite{ZTBZ09} to characterize the afterpulsing distribution
in a 1\,$\mu$s range, see Figure \ref{dis}.
The laser repetition frequency is as low as 500\,kHz and the coincidence window
is 50\,ns. The result is shown in the histogram of Figure \ref{dis},
in which the constant dark count contribution has been subtracted.
$P_{ap}$[ns$^{-1}$] is also shown in the right axis of Figure \ref{dis} and calculated as
\begin{equation}
\label{}
P_{ap}[ns]=\frac {P_{ap}}{50\cdot t_{g}\cdot f_{g}},
\end{equation}
where $t_{g}\cdot f_{g}$ is the duty cycle and at each bin the minor
afterpulsing contribution from the previous bins is corrected through the iterative calculations.
The curve is fitted using a multiple detrapping model \cite{ZTGGZ09}, which
suggests that there are mainly two kinds of detrapping types with
a quite short lifetime of $\sim$100\,ns and a relatively long lifetime of $>$1\,$\mu$s. In Figure \ref{dis}, $P_{ap}$[ns$^{-1}$]
reaches 5$\times10^{-5}$ ns$^{-1}$, equivalent to 10$\cdot P_{dc}$[ns$^{-1}$], at the time of 900\,ns. For comparison, in the case of integrated
active quenching system \cite{ZTGGZ09}, the typical value with the same
temperature and time settings is 7$\times10^{-3}$ ns$^{-1}$, equivalent to 2058$\cdot P_{dc}$[ns$^{-1}$] ($P_{dc}$[ns$^{-1}$]=3.4$\times10^{-6}$ ns$^{-1}$),
and a deadtime setting $>$ 20\,$\mu$s is required to suppress $P_{ap}$[ns$^{-1}$] down to 5$\times10^{-5}$ ns$^{-1}$,
equivalent to 14.7$\cdot P_{dc}$[ns$^{-1}$], see Figure 8(b) in Ref. \cite{ZTGGZ09}.

\section{QKD SIMULATIONS}
\label{sec:qkd}

\begin{figure}[t]
\centering
\includegraphics[width=12 cm]{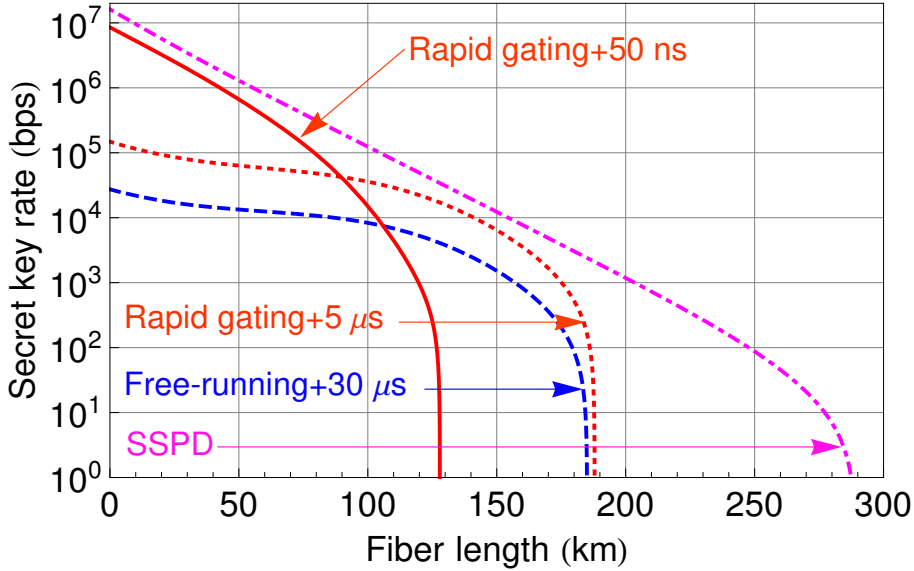}\\
\caption{The simulation of secure key rate for the COW QKD protocol
versus fiber distance, using the rapid gating SPAD with 50\,ns (solid)
and 5\,$\mu$s (dotted) deadtime and the free-running SPAD \cite{TSGZR07,ZTGGZ09} with 30\,$\mu$s deadtime (dashed),
as well as SSPD (dot-dashed), respectively.}
\label{cowrapid}
\end{figure}

Rapid gating detection would appear to be well suited to high-speed QKD applications \cite{NFIHT07,DYDSS08}.
To look more closely at this, we model and simulate the secure key rate for the coherent one-way (COW) QKD \cite{COW}
scheme, as a function of distance, based on our rapid gating and free-running SPADs \cite{TSGZR07,ZTGGZ09}
as well as SSPDs \cite{SSPD}, see Figure \ref{cowrapid}.
The typical parameters of the QKD system for modeling such as mean photon number per pulse,
decoy pulse probability, insertion loss for Bob's system, interferometric visibility etc.,
are taken from Ref. \cite{COW}. The typical parameter values of the rapid gating
and free-running SPADs at 10\,\% efficiency are taken from this experiment and Ref. \cite{COW}, respectively.
For a fair comparison, the parameters of the SSPD are assumed as follows, 10\,\% system detection
efficiency, 10\,Hz dark count rate and minimum pulse width of 20\,ns, corresponding to
maximum count rate of 50\,MHz. These SSPD parameters are probably not the best results reported
so far, but they are not necessarily underestimated. For instance,
the system detection efficiency of the SSPD in our group is 2.6\,\% with 10\,Hz dark count rate \cite{250}.
We also assume that the detection window is 100\,ps for all the three kinds of detectors.

In the rapid gating case, the crucial parameter, the afterpulse probability is estimated as
\begin{equation}
\label{int}
\overline{P_{ap}}=\int_{\tau_{d}}^{\overline{\Delta T}}f(t)dt,
\end{equation}
where $\overline{P_{ap}}$ is the average afterpulse probability, $\tau_{d}$ is the deadtime,
$\overline{\Delta T}$ ($>\tau_{d}$) is the average time interval between two detections,
and $f(t)$ is the above mentioned fitting function in Figure \ref{dis}. We focus on two
extreme regimes: short distances with high rates and maximum distances.

Firstly, let us look at short distances. The rapid gating scheme
with $\tau_{d}$=50\,ns provides considerably high rates over short
distances, e.g., Mbps key rates for distances less than 40\,km. This results are much better
than the free-running SPAD and approach the results of SSPD, which implies that the rapid gating
SPAD with small $\tau_{d}$ is well suited in short distance regimes.

Secondly, the maximum distance in this case is limited to $\sim$130\,km.
However, if $\tau_{d}$ is increased, the maximum distance can be increased accordingly
because for longer distances the probability of a photon arriving is significantly
reduced and hence we can increase $\tau_{d}$ to suppress the afterpulsing without
adversely affecting the rates. When $\tau_{d}$=5\,$\mu$s, the key rates
are reduced for short distances compared to $\tau_{d}$=50\,ns but the
maximum distance is extended up to $\sim$190\,km, see the dotted
line in Figure \ref{cowrapid}. Moreover, we find that we approach the
distance limit for the ideal case with the assumption of no
afterpulsing, which implies that the afterpulse probability of
rapid gating SPAD can be suppressed down to approximately zero with $\tau_{d}$=5\,$\mu$s.
Interestingly, the free-running SPAD with $\tau_{d}$=30\,$\mu$s can implement almost the
same maximum distance, probably because $\tau_{d}$=30\,$\mu$s is sufficient
to suppress the afterpulse probability of free-running SPAD down to a negligible level and
therefore the maximum distance is only limited by the dark count characteristics.

Finally, we model the system with SSPD for comparison,
see the dot-dashed line in Figure \ref{cowrapid}.
We see that for short distances, SSPDs have only a minimal advantage
over rapid gating SPADs in terms of rate.
Nevertheless, the maximum distance that SSPDs can obtain is more than 280\,km,
due to the ultralow noise characteristics of such detectors. This
suggests that SSPDs are well suited for ultra-long distance ($>$200\,km) applications,
which was already verified by some previous experiments \cite{40db,250,pan}. The maximum distance difference
between the rapid gating SPADs and the SSPDs is essentially due to the difference
in dark count characteristics. For instance, the dark count rate of our rapid gating SPAD is 480 times
higher than that of SSPD. There is still some room to optimize the rapid 
gating SPAD, e.g., cooling down further the temperature of the SPAD while 
increasing the deadtime setting. However, short distances with high key rates
are more interesting than ultra-long distances with ultralow key rates for practical
applications. 

In general, we can conclude that the afterpulsing is no longer a limiting factor for QKD.
For high-rate QKD applications, rapid gating SPADs with small $\tau_{d}$ are favorable
candidates. Both free-running and rapid gating SPADs are well suited for long-distance
applications, say, $<$200\,km, while SSPDs remain advantageous for
distances $>$200\,km. In practice, rapid gating SPADs are definitely the
appropriate choice compared to SSPDs, due to the disadvantages of SSPDs
such as cryogenic requirements and non-cost-effectiveness.

\section{CONCLUSIONS}
\label{sec:con}

In conclusion, we demonstrate a near-infrared single-photon detector, based on an
InGaAs/InP SPAD, capable of synchronized operation at 2.23\,GHz clock rates.
This scheme can effectively suppress the afterpulsing. We illustrate a
useful technique to characterize the afterpulsing distribution for easy
comparison between different systems. We also demonstrate the performance
impact of these types of devices on QKD. Rapid gating SPADs are well
suited to both high-rate and long-distance QKD applications, and modeling suggests
that the maximum distances can reach $\sim$190\,km. Most importantly,
we conclude that the afterpulsing is no longer a limiting factor for QKD.
Finally, dark count remains the most of important bottleneck for further increases in
the maximum achievable distances. Suppressing dark count rates of SPADs is still an open
issue that deserves investigation to further extend QKD distances with practical detection schemes.

\acknowledgments
The authors thank O. Guinnard and N. Gisin for useful discussions and
acknowledge financial support from the Swiss NCCR-Quantum Photonics.



\begin{thebibliography}{99}
\bibitem{GRTZ02}
Gisin, N., Ribordy, G., Tittel, W. and Zbinden, H.,
``Quantum cryptography,''
Rev. Mod. Phys. 74, 145 (2002).

\bibitem{ELZZG10}
Eraerds, P., Legre, M., Zhang, J., Zbinden, H. and Gisin, N.,
``Photon Counting OTDR: Advantages and Limitations,''
arXiv:1001.0694v2.

\bibitem{RGZG98}
Ribordy, G., Gautier, J.D., Zbinden, H. and Gisin, N.,
``Performance of InGaAs/InP Avalanche Photodiodes as Gated-Mode Photon Counters,''
Appl. Opt. 37, 2272 (1998).

\bibitem{Itzler07}
Itzler, M. A., Ben-Michael, R., Hsu, C.-F., Slomkowski, K., Tosi, A., Cova, S., Zappa, F. and Ispasoiu, R.,
``Single photon avalanche diodes (SPADs) for 1.5 $\mu$m photon counting applications,''
J. Mod. Opt. 54, 283 (2007).

\bibitem{Cova96}
Cova, S., Ghioni, M., Lacaita, A., Samori, C. and Zappa, F.,
``Avalanche photodiodes and quenching circuits for single-photon detection,''
Appl. Opt. 35, 1956 (1996).

\bibitem{TSGZR07}
Thew, R. T., Stucki, D., Gautier, J.-D., Zbinden, H. and A. Rochas,
``Free-running InGaAs/InP avalanche photodiode with active quenching for single photon counting at telecom wavelengths,''
Appl. Phys. Lett. 91, 201114 (2007).

\bibitem{ZTGGZ09}
Zhang, J., Thew, R., Gautier, J.-D., Gisin, N. and Zbinden, H.,
``Comprehensive Characterization of InGaAs¨CInP Avalanche Photodiodes at 1550 nm With an Active Quenching ASIC,''
IEEE J. Quantum Electron. 45, 792 (2009).

\bibitem{NSI06}
Namekata, N., Sasamori, S. and Inoue, S.,
``800 MHz single-photon detection at 1550-nm using an InGaAs/InP avalanche photodiode operated with a sine wave gating,''
Opt. Express 14, 10043 (2006).

\bibitem{NAI09}
Namekata, N., Adachi, S. and Inoue, S.,
``1.5 GHz single-photon detection at telecommunication wavelengths using sinusoidally gated InGaAs/InP avalanche photodiode,''
Opt. Express 17, 6275 (2009).

\bibitem{YKSS07}
Yuan, Z. L., Kardynal, B. E., Sharpe, A. W. and Shields, A. J.,
``High speed single photon detection in the near infrared,''
Appl. Phys. Lett. 91, 041114 (2007).

\bibitem{DDYSBS09}
Dixon, A. R., Dynes, J. F., Yuan, Z. L., Sharpe, A. W., Bennett, A. J. and Shields, A. J.,
``Ultrashort dead time of photon-counting InGaAs avalanche photodiodes,''
Appl. Phys. Lett. 94, 231113 (2009).

\bibitem{ZTBZ09}
Zhang, J., Thew, R., Barreiro, C. and Zbinden, H.,
``Practical fast gate rate InGaAs/InP single-photon avalanche photodiodes,''
Appl. Phys. Lett. 95, 091103 (2009).

\bibitem{NFIHT07}
Namekata, N., Fujii, G., Inoue, S., Honjo, T. and Takesue, H.,
``Differential phase shift quantum key distribution using single-photon detectors based on a sinusoidally gated InGaAs/InP avalanche photodiode,''
Appl. Phys. Lett. 91, 011112 (2007).

\bibitem{DYDSS08}
Dixon, A. R., Yuan, Z. L., Dynes, J. F., Sharpe, A. W. and Shields, A. J.,
``Gigahertz decoy quantum key distribution with 1 Mbit/s secure key rate,''
Opt. Express 16, 18790 (2008).

\bibitem{COW}
Stucki, D., Barreiro, C., Fasel, S., Gautier, J., Gay, O., Gisin, N., Thew, R.,
Thoma, Y., Trinkler, P., Vannel, F. and Zbinden, H.,
``Continuous high speed coherent one-way quantum key distribution,''
Opt. Express  17, 13326 (2009).

\bibitem{SSPD}
Gol'tsman, G. N., Okunev, O., Chulkova, G., Lipatov, A., Semenov, A., Smirnov, K., Voronov, B.,
Dzardanov, A., Williams, C. and Sobolewski, R.,
``Picosecond superconducting single-photon optical detector,''
Appl. Phys. Lett. 79, 705 (2001).

\bibitem{40db}
Takesue, H., Nam, S. W., Zhang, Q., Hadfield, R. H., Honjo, T., Tamaki, K. and Yamamoto, Y.,
``Quantum key distribution over a 40-dB channel loss using superconducting single-photon detectors,''
Nat. Photonics 1, 343 (2007).

\bibitem{250}
Stucki, D., Walenta, N., Vannel, F., Thew, R. T., Gisin, N., Zbinden, H., Gray, S.,
Towery, C. R. and Ten, S.,
``High rate, long-distance quantum key distribution over 250 km of ultra low loss fibres,''
New Journal of Physics 11, 075003 (2009).

\bibitem{pan}
Chen, T. Y., Wang, J., Liu, Y., Cai, W. Q., Wan, X., Chen, L. K., Wang, J. H., Liu, S. B., Liang, H., Yang, L., Peng, C. Z., Chen, Z. B. and Pan, J. W.,
``200 km Decoy-state quantum key distribution with photon polarization,''
arXiv:0908.4063v1.

\end{thebibliography}

\end{document}